# ROBO-AO M DWARF MULTIPLICITY SURVEY: Catalog

Claire Lamman,[1,2,3] Christoph Baranec,[2] Zachory K. Berta-Thompson,[1] Nicholas M. Law,[4]
Jessica Schonhut-Stasik,[2] Carl Ziegler,[5] Maïssa Salama,[2] Rebecca Jensen-Clem,[6] Dmitry A. Duev,[7]
Reed Riddle,[7] Shrinivas R. Kulkarni,[7] Jennifer G. Winters,[3] and Jonathan M. Irwin[3]

[1]Department of Astrophysical and Planetary Sciences, University of Colorado, Boulder, CO 80309, USA
[2]Institute for Astronomy, University of Hawai'i at Mānoa, Hilo, HI 96720-2700, USA
[3]Harvard-Smithsonian Center for Astrophysics, 60 Garden St., Cambridge, MA 02138, USA
[4]Department of Physics and Astronomy, University of North Carolina at Chapel Hill, Chapel Hill, NC 27599-3255, USA
[5]Dunlap Institute for Astronomy and Astrophysics, University of Toronto, Ontario M5S 3H4, Canada
[6]Astronomy Department, University of California, Berkeley, CA 94720, USA
[7]Division of Physics, Mathematics, and Astronomy, California Institute of Technology, Pasadena, CA 91125, USA

## ABSTRACT

We analyze observations from Robo-AO's field M dwarf survey taken on the 2.1m Kitt Peak telescope and perform a multiplicity comparison with Gaia DR2. Through its laser-guided, automated system, the Robo-AO instrument has yielded the largest adaptive optics M dwarf multiplicity survey to date. After developing an interface to visually identify and locate stellar companions, we selected eleven low-significance Robo-AO detections for follow-up on the Keck II telescope using NIRC2. In the Robo-AO survey we find 553 candidate companions within 4″ around 534 stars out of 5566 unique targets, most of which are new discoveries. Using a position cross match with DR2 on all targets, we assess the binary recoverability of Gaia DR2 and compare the properties of multiples resolved by both Robo-AO and Gaia. The catalog of nearby M dwarf systems and their basic properties presented here can assist other surveys which observe these stars, such as the NASA TESS mission.

## 1. INTRODUCTION

M dwarfs account for nearly three-quarters of stars in our solar neighborhood (Henry et al. 2006), yet there's still much to learn about which nearby M dwarfs host stellar companions and how these small stars form. Finding and characterizing M dwarf multiples is useful for studying transiting exoplanets, and multiplicity trends among them can yield insight into stellar formation and evolution.

M dwarfs are favorable targets for transiting exoplanet surveys since they and their planets are abundant (Dressing & Charbonneau 2015), tend to have planetary systems which are relatively compact (Muirhead et al. 2015), and a planet will have a deeper transit depth around an M dwarf than a larger star (Charbonneau & Deming 2007). The Transiting Exoplanet Survey Satellite (TESS) is targeting main-sequence dwarf stars with spectral types M5-F7 (Ricker et al. 2014). Current expectations for the number of planets that TESS will discover around M dwarfs range from 500-1000 (Barclay et al. 2018; Ballard 2019). However, binaries can produce false planet transits and a planet radius can be significantly underestimated if its host has an unknown stellar companion (Ciardi et al. 2015; Ziegler et al. 2018a). M dwarf binary separations peak below 10AU (Gizis et al. 2003; Burgasser et al. 2007; Duchene & Kraus 2013; Ward-Duong et al. 2015), which corresponds to an angular separation less than 1″ for objects beyond 10pc. This is significantly less than the TESS pixel size of 21″ (Ricker et al. 2014), and ground-based transit follow-up observations are typically limited to 1″ seeing (Collins et al. 2018). Therefore, it is important to use high-angular resolution surveys to identify and characterize sub-arcsecond binaries.

Once systems are identified, overall multiplicity patterns can reveal properties ubiquitous to star formation.

Corresponding author: Claire Lamman
claire.lamman@cfa.harvard.edu

Useful statistics include the total frequency of multiples, the distribution of physical separations, mass ratios, and how these characteristics change as a function of primary mass. Empirically determining these statistics provides a check of star formation models and is especially useful to the study of stellar multiples. For instance, orbital period distribution can distinguish between either a spatial scale or scale-free formation process for binaries, and orbital eccentricity is a key characteristic of how systems evolve (Duchene & Kraus 2013). Additionally, since the M spectral classification spans a factor of eight in mass, it is unclear if these stars even belong in the same homogeneous population. Solar-type stars consistently display different multiplicity properties than brown dwarfs (Burgasser et al. 2007). G type stars and their companions have an orbital period distribution that peaks around 300 years (corresponding to 51 AU), a roughly uniform distribution in mass ratios between 0.2-0.95, and a multiplicity rate of 46% (Raghavan et al. 2010). Brown dwarf systems have typical periods of <40 years, mass ratios that rise toward near-equal masses, and a multiplicity of 10-30% (Burgasser et al. 2007). As the intermediate between these two populations, understanding the multiplicity of M dwarfs will help reveal if there is a common formation process along the lower main sequence or two distinct ones for low and high mass populations (Bate 2012). Most of these statistics require an extensive M dwarf multiplicity survey, a technical challenge that has only become possible relatively recently.

Outside of our immediate solar neighborhood (Henry et al. 2006), large-scale M dwarf multiplicity surveys are challenging due to the high angular resolution required to resolve the faint, typically close companions. Several early surveys, which employed radial velocity, direct imaging of close systems, and speckle interferometry, were combined and analyzed by Fischer & Marcy (1992). They estimated an overall multiplicity of $42 \pm 9\%$, which was consistent with solar-type stars; however, they found a peak in the separation distribution at 3-30AU, which is at smaller separations than that of solar-type stars. Later studies obtained a more complete survey of a volume-limited sample. A Hubble survey (Gizis et al. 2003) of late M dwarfs determined that physical separation instead peaks around 2-4 AU, significantly different from the solar-type population. A survey combining adaptive optics, infrared interferometric data, and radial velocity (Delfosse et al. 2004) found a separation distribution similar to solar-type stars within 10 AU but not beyond. They also saw a flat mass ratio distribution for periods above 50 days but a clear tendency toward equal masses for shorter periods, and an overall multiplicity of $26 \pm 3\%$. More recent multiplicity surveys have used Lucky Imaging: a technique of using only the highest-quality fraction of many short exposures. These include Law et al. (2010), which determined a multiplicity of $13.6^{+6.5}_{-4}\%$ for late-type M dwarfs, Bergfors et al. (2010), which a determined multiplicity of $32 \pm 6\%$ with a flat mass distribution for both early and late M dwarfs, and Janson et al. (2012), which determined a multiplicity of 34.4% and confirmed a uniform mass ratio distribution. The largest of these Lucky-Imaging surveys (Janson et al. 2012), observed 701 M dwarfs and found 205 systems. These studies were followed by several which utilized pre-existing large-scale surveys, such as (Shan et al. 2015) which found 12 eclipsing binaries out of 3905 stars in the Kepler field, and (Ward-Duong et al. 2015) which combined AO with Sloan plates to find a multiplicity of $28.6^{+2.7}_{-3.1}\%$ and a separation peak at 5.9 AU. One of the largest M dwarf multiplicity surveys, SLoWPoKES, found 1342 widely-separated ($\gtrsim 500$AU) proper motion pairs through another Sloan archive search (Dhital et al. 2010). However, the SLoWPoKES multiplicity rate of 1.1% represents less than 4% of M dwarf companions when compared to the expected total multiplicity rate. Currently, the most comprehensive M dwarf multiplicity study, Winters et al. (2019), surveyed a volume-limited sample of 1120 stars through new observations, archival data, and a thorough literature search. They found a multiplicity of $26.8 \pm 1.4\%$ and a distribution peak between 4-20 AU.

Based on these results, we expect that roughly one third of the M dwarf stars targeted by TESS will be multiples. Therefore, it's important to identify and characterize more of these systems. To do this, we have used Robo-AO: an autonomous laser adaptive optics system that can achieve near diffraction-limited imaging without Lucky Imaging, and observe ~150 targets each night (Jensen-Clem et al. 2018). Here we focus on the catalog creation and a comparison to Gaia DR2, leaving a statistical analysis of overall multiplicity rates and trends for a future work. This paper is structured as follows: in Section 2 we describe our target selection and observations. We explain the data reduction process in Section 3. In Section 4 we combine and compare our results with Gaia DR2 to assess DR2 binary recoverability and obtain estimates of physical system properties. This is followed by a discussion of biases within our sample and the potential application of results in Section 5, then we conclude and discuss future work with Section 6. Tables for the systems and targets observed are included in the Appendix.

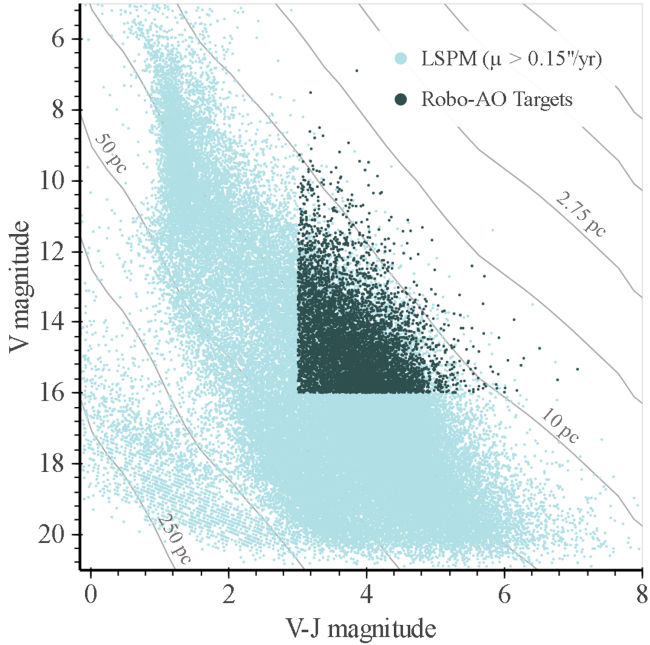

**Figure 1.** Our color selection from LSPM with lines of constant photometric distance. These distances are estimated from empirical magnitude-distance relations (Pecaut & Mamajek 2013) and are based on the assumption that these are all single, main-sequence stars. As high-proper motion, red objects, the majority of these are likely nearby, northern M dwarfs.

## 2. TARGET SELECTION AND OBSERVATIONS

We selected nearby red dwarfs and observed each target with Robo-AO. Follow-up observations were later done on select targets with Keck. This section presents the details of the target selection and these observations.

### 2.1. *Target Selection*

To focus on nearby stars, we selected 7,083 northern ($dec > 0°$) targets from the Lépine and Shara Proper Motion (LSPM) catalog (Lépine & Shara 2005). Every star in this catalog has a high proper motion ($> 0''15/yr$); nearby objects generally move faster on the sky than background stars, so a high proper motion is typically indicative of a close star. To ensure that our sample of close stars was also apparently bright and red, we then made cuts of $V < 16$ mag and $V - J > 3$ mag, using the given LSPM values (Fig 1). These cuts approximately correspond to masses less than $0.6 M_\odot$ and spectral types cooler than M0 for typical, main-sequence stars within 50pc (Pecaut & Mamajek 2013). Although these criteria are designed to select only intrinsically faint, red stars, seven objects were brighter than expected for dwarfs ($< 6V$ mag), and we confirmed all to be known red giants through SIMBAD (Wenger et al. 2000). These were discarded from analysis. We expect less than 45 of the remaining objects to be background giants based on cuts presented in (Lépine & Gaidos 2011). Estimated LSPM photometric distances initially placed the majority of our targets within 30pc, which often turned out to be a significant underestimate based on Gaia DR2 parallax measurements (Section 4).

### 2.2. *Robo-AO*

We obtained 7046 high-angular-resolution images of 6793 unique targets over the course of 211 nights between 2015 December 18 through 2017 June 8. Out of the original sample, 4% (290) of the targets were not observed with the intelligent queue. Each image was taken in the $i'$-band with a 90s exposure time. The observations were performed using the Robo-AO laser adaptive optics system (Baranec et al. 2014) mounted on the Kitt Peak 2.1-m telescope, masked to a 1.85-m aperture. The adaptive optics system runs at a loop rate of 1.2 kHz to correct high-order wavefront aberrations and images are recorded on an electron-multiplying CCD (EMCCD) at 8.6Hz which allowed for post-facto image displacement correction in software using the target as a natural guide star. The median seeing at the 2.1-m telescope was $1''44$ which resulted in an average $i'$-band Strehl ratio of 4% and a full-width at half-maximum of $\sim 0''12$ (Jensen-Clem et al. 2018). Specifications of the Robo-AO M dwarf survey are summarized in Table 1.

**Table 1.** Robo-AO M dwarf survey specifications

| | |
|---|---|
| Number of targets | 6793 |
| Telescope | Kitt Peak 2.1m telescope |
| Camera | Andor iXon DU-888 |
| Observation Wavelength | $i'$ |
| Exposure time | 90s |
| Field of view | $36'' \times 36''$ |
| Pixel scale | 35.1 mas/pixel |
| Detector format | $1024^2$ pixels |
| Observation date range | 2015 December 18 - 2017 June 8 |

### 2.3. *Automated Imaging Pipeline*

At the end of each observing night, the Robo-AO system processes and archives observations via automatic pipelines, detailed in Jensen-Clem et al. (2018). To summarize, data is first shifted and added in a process that is optimized for either high or low signal-to-noise images. Then, to better detect faint companions, a "high-contrast imaging pipeline" creates a custom, locally optimized PSF-subtracted image of a $3''5$ cut-out around the target. The PSF is generated from a reference library of single-star Robo-AO observations from the same night. They are combined via Karhunen-Loéve image processing to create a synthetic PSF. After a high-pass filter is applied to the image, we then subtract the generated PSF image. The pipeline also creates a contrast curve for each observation. This is five times the standard deviation of the noise measured radially outward and normalized by the stellar flux; it represents the faintest contrast that could likely be detected at $5\sigma$ for a given separation. (Jensen-Clem et al. 2018).

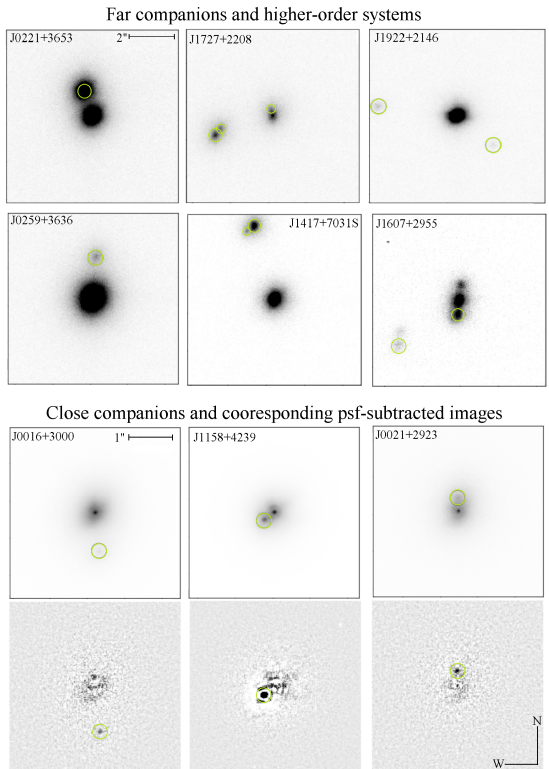

**Figure 2.** Robo-AO observations of nine candidate systems; companions are circles in green. LSPM J1607+2955 is both a triple system and an example of a falsely-tripled image. LSPM J1727+2208 is a quadruple: the highest-order system we found.

### 2.4. *Additional Imaging with Keck*

After initial data analysis, we obtained additional imaging of 10 potentially very close (separations$< 0\rlap{.}''3$) companions and a potential triple system using the NIRC2 imager behind laser-guidestar adaptive optics on the 10m Keck II telescope on 2017 August 3. These targets were chosen from 30 Robo-AO targets which potentially had close companions causing a slight false-tripling effect (Section 3.2), but could not be visually confirmed. The 20 for which we could not obtain follow-up observations due to sky position and available observing time were discarded as multiples. The observations' pixel scale is 9.9 mas/pixel, with a $10''$ field of view. For targets with a visible companion in the first image taken and displayed with the NIRC2 GUI, we took multiple images in the $J$ filter and in $K$ or $Kp$. Observations specifications for each target, including exposure times, can be found Table 2 in the appendix. These images were sky-subtracted using the median pixel value of all images taken of the same target, in the same filter. They were then flat-field calibrated and stacked to create a final image for each filter.Further analysis for companion parameters is detailed in Section 3.3.

## 3. IDENTIFYING MULTIPLES

We developed an interface to perform several visual checks on each target. This provided us with the relative position of companions which we used to obtain contrast ratios for each multiple.

### 3.1. *Visual Inspection*

After passing through the automated pipeline, each observation needed to be visually inspected to ensure data quality, correct telescope pointing, and that the correct target in the field was reduced. The identification and locating of companions also needed to be done manually. To efficiently examine all observations, we developed and used a Graphical user interface for Robo-AO M dwarfs (`GRAM`).

This program was created using the Python module tKinter (Shipman 2013) and made use of contrast curves and PSF-subtracted images from the automated pipeline, a STScI Digitized Sky Survey database image, and scaled images of the original stacked FITS file. `GRAM` allows a user to assign a variety of tags to each observation: "good", "bad", "uncertain", "not enough stars", "incorrect pointing", "needs different database", "possible close binary", and "needs manual inspection". The user can also select a corrected location of the target, if needed, and the location of a potential companion in either a view of $8''$ centered around the primary star or a set of images displaying a 1.5$''$ view centered around the primary, where is it easier to visually identify closer companions (Fig 3). This returns a text file of each observation's information which can be compiled into an ensemble file. `GRAM` can be applied to other surveys requiring large-scale visual inspections, and was successfully tested by re-analyzing data from the Robo-AO Kepler Asteroseismic survey (Schonhut-Stasik et al. in submission). Using `GRAM`, we confirmed all multiples found in the original manual search and discovered two new companions. The source code for `GRAM` is publicly available (Lamman 2019).

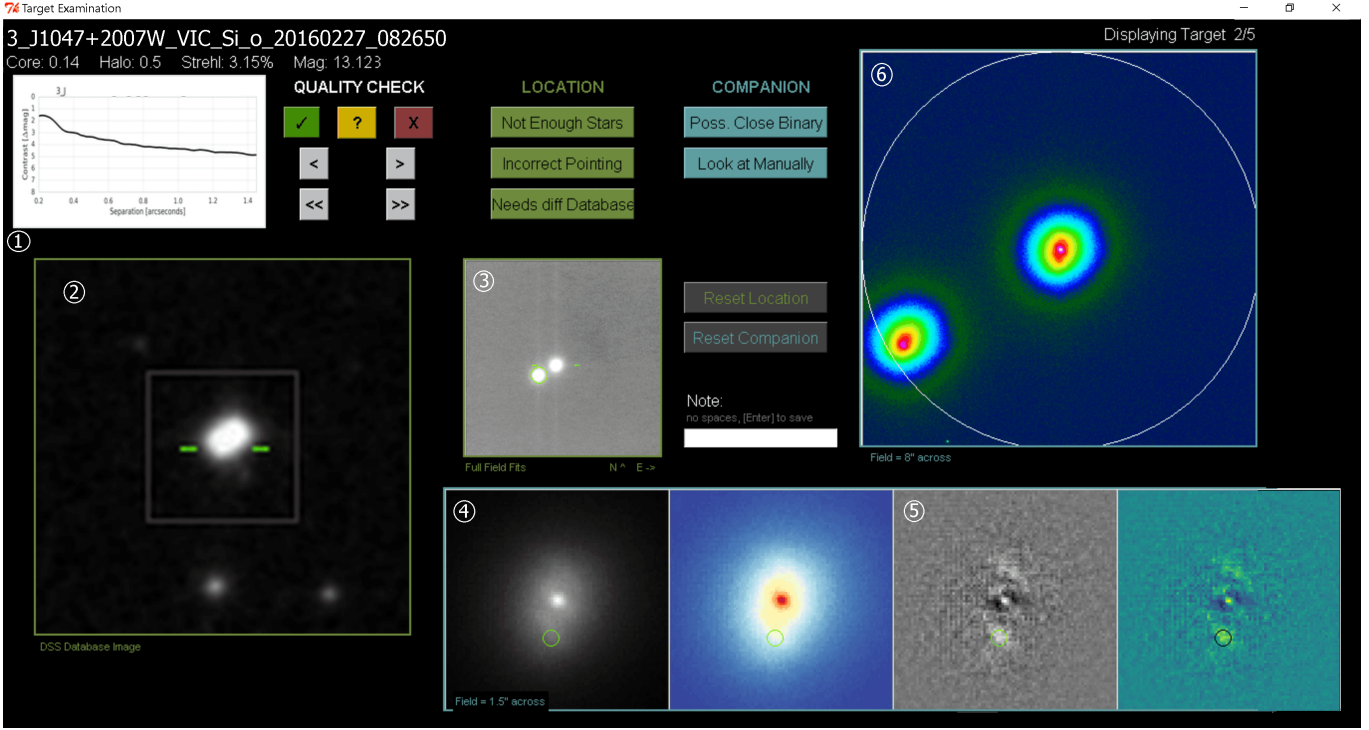

**Figure 3.** `GRAM` screen shot of a candidate triple system. In this example, the pipeline reduced the wrong star, it is a quality observation, and there is a potential close and far companion. This observation also has a slight false triple effect about 0.″25 above the primary star.

Quality Check gives the user the option to change the default "good" tag, and below are controls to progress through the targets either one or ten at a time. The blue and green rectangular buttons add related tags. Resetting the location or companion changes the tags back to default and removes any positions that may have been selected. ① A contrast curve for the observation created by the automatic pipeline (Section 2.3). ② STScI Digitized Sky Survey image with a 35.″6 square (the Robo-AO field of view) and target location marked in green. ③ The full-field Robo-AO image with the star that has been reduced by the automatic pipeline marked by green lines. If the wrong object is marked, the user can click on the correct star. This circles and saves the corrected location. ④ Two cut-outs of the stacked pipeline image showing 0.″75 around the primary star next to ⑤ two pipeline PSF-subtracted 0.″75 cut-outs. These four images are for identifying close companions. Clicking in any image circles the selected location in each view and saves the potential companion's pixel coordinates. ⑥ The 8″ frame, displaying a radius of 4″ around the primary star. This is the maximum separation we are searching for. Clicking here also marks and saves the selected location.

We analyzed the observations through `GRAM` in batches of varying quality based on core size. This is the width, in arcseconds, of the star's PSF core and is a proxy for image performance (Law et al. 2014). After visually examining all targets, our final sample contained 5566 unique targets which were marked as having high-quality observations in which the automatic pipeline had also reduced the correct target. 581 observations were marked as having one or more potential companions, of which nine had two nearby stars and one had three (Fig 2).

### 3.2. Companion Locations and Contrast Ratios

The location of each secondary star is given by its position angle and apparent distance from the primary. To determine the primary's pixel coordinates, we used a local box-centroid with a 3-pixel radius for systems with separations smaller than 0″.75, and a 5-pixel radius for those farther. Since both the primary flux and stacking procedure (Section 2.3) affected the secondary psf, we used the location selected manually in GRAM for the secondary's position. Positions were also adjusted to account for Robo-AO's slight distortion pattern, which is expressed in (Jensen-Clem et al. 2018). The separation and position angle errors are estimated by assuming a maximum pixel uncertainty of 2 for both stars' positions.

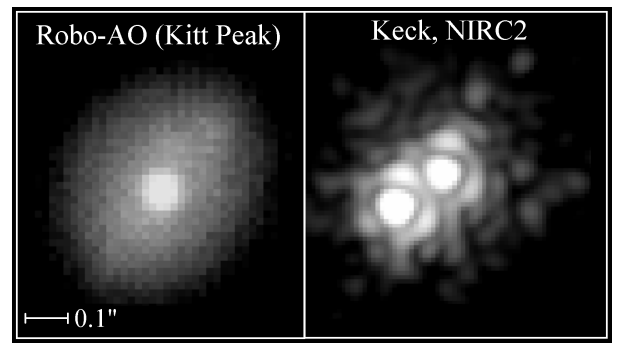

**Figure 4.** Robo-AO discovery image of a companion to LSPM J1648+1038 taken in $i'$ compared to the same system as seem by NIRC2 in Kp; each image is scaled individually. This is one of our closest systems at an angular separation of 0.″135±0.02. It was identified in the Robo-AO data from its PSF-subtracted image and the oblong shape created by the false-tripling effect often seen in close companions.

Contrast ratios are the difference between the magnitude of the primary and secondary star. We used Photutils aperture photometry (Bradley et al. 2018) to obtain flux estimates for each star by summing the counts within an aperture around the star and subtracting the expected background for an equal area (Fig 5). The background was determined independently for each star by taking the average of four apertures. These background apertures were located on the opposite side of the other star, at the same distance away from it as the star being measured (90−270° away from the star's position around its companion). This helps account for the other star's flux by including it in the background determination. To ensure that the background apertures didn't overlap or contain any flux from the star being measured, we varied the aperture radius. The radius is 15 pixels for all systems with a separation greater than 1″, and linearly decreases with separation to 4 pixels at 0″.1. This may create a systematic effect on the contrasts of close companions. Contrast errors were estimated by summing the standard deviation of the background aperture counts in quadrature with both stars' photon-counting errors. 28 companions with a counting error larger than $0.25\Delta mag$ or a background error which contributed more than $1\Delta mag$ to the total contrast error were removed from further analysis. The errors for most close companions ($< 0″.5$) are still larger than $0.13\Delta mag$, which corresponds to a $SNR < 7$. By choosing to include these, we believe a companion exists but that there is a large systematic error from our photometry at close separations. We encourage re-observing these systems for more confident contrast ratios.

The automatic pipeline stacks images relative to the brightest pixel in the frame. For some low-contrast companions (similar brightness to the primary), images will be stacked with an inconsistent orientation, causing part of the companion's flux to be mirrored on the opposite side of the primary star. This is the "false tripling" effect, of which an example is shown in Figure 2. We obtained contrast ratios for these 106 cases by changing the secondary star's background aperture locations to 60, 90, 240, and 300° around the primary to avoid the false triple. We performed the same photometry on the false triple and then determined a final contrast via the process outlined in Law (2006). Although the true companion location is often significantly brighter than the false companion, it can be ambiguous and may result in an 180° phase difference.

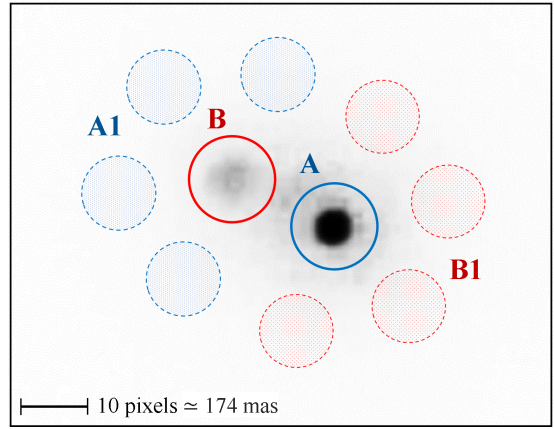

**Figure 5.** Aperture photometry example for LSPM J1219+0214, a close (0″.30 separation) companion. A: aperture around primary star. B: aperture around secondary star. A1: four apertures on the opposite side of the secondary. The average of these is subtracted from the count within A. B1: four apertures on the opposite side of the primary. The average of these is subtracted from the count within B. As a reduced image, the pixel scale here is different than the Robo-AO plate scale of 35.1 mas/pixel.

Our final companion locations and contrast ratios are displayed in Fig 6 and can be found in Tables 3 and 4, where the false triple cases are marked.

### 3.3. Confirming with Keck

The goal of this analysis is to determine more reliable contrast ratios and separations for the very few targets that were observed with Keck. We obtained follow-up observations of 11 targets with potential companions using the NIRC2 imager on Keck II (Section 2.4). Seven companions were confirmed, including those in the triple system LSPM J1606+0823 and one of our closest detected companions at $0.135\pm0″.02$ (Fig 4). From these we obtained updated separations and contrast ratios, which can be found in Appendix Table 2. Using the final image resulting from the process in Section 2.4, we determined separations and contrast ratios using the same Photutils photometry described in Section 3.2, except with centroid fitting for the location of both stars and by applying the distortion solution from Service et al. (2016).

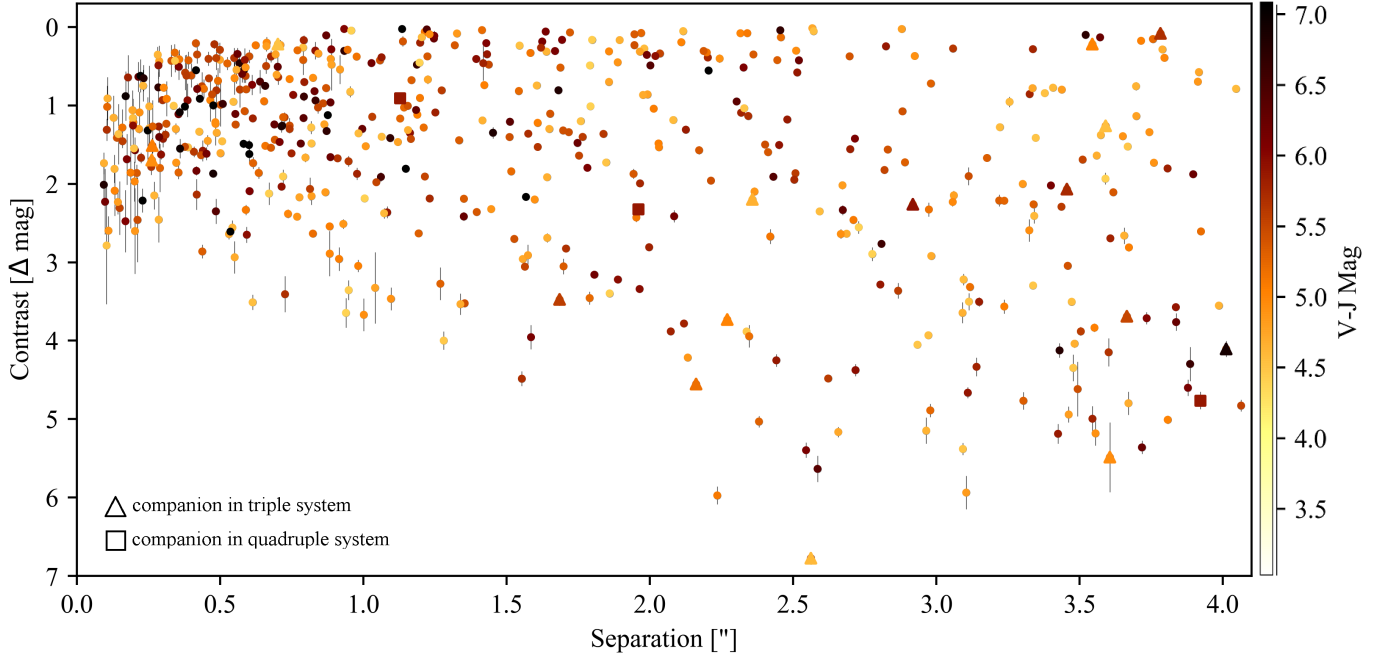

**Figure 6.** Contrast ratios and angular separations for 553 Robo-AO companions. The closer two stars are and the higher their contrast, the more difficult they are to resolve. Almost all companions within $0\rlap{.}''5$ are low-significance ($< 8\sigma$) detections (Tables 3, 4) and require further observations for reliable photometry. Points are colored based on the LSPM target's estimated V-J magnitude.

## 4. GAIA DR2 CROSS MATCH AND PHYSICAL PROPERTIES

Here we compare and combine our results with Gaia's second data release (DR2), which we use to assess the physical association of companions and explore their basic properties.

### 4.1. Cross Match

We matched 6214 of our original 6793 LSPM targets with Gaia DR2 identifiers (Gaia Collaboration et al. 2016; Brown et al. 2018). This was done by selecting the Gaia object for each LSPM target with the lowest $d$ value as long as $d < 5$, where:

$$d = \left(\frac{\Delta position}{1''}\right)^2 + \left(\frac{\Delta pm}{0.1''/yr}\right)^2 \qquad (1)$$

$\Delta position$ is the distance in arcseconds between the location of the LSPM star and the Gaia star's projected position, after using the LSPM proper motion to project each from its catalog 2000.0 epoch to Gaia DR2's 2015.5 epoch. $\Delta pm$ is the proper motion difference between the LSPM and Gaia DR2 catalog values, in $arcsec/yr$. This value is weighted to select the physically closest object in most cases, but also takes into account the possibility that one of these high-proper motion stars is passing by another object. 5730 of these LSPM objects are already matched to DR2 identifiers on the

Gaia archive[1] through their 2MASS IDs (Skrutskie et al. 2006). Our cross match agreed with all but four: LSPM J1953+1136, LSPM J2336+3939, LSPM J1938+2127, and LSPM J0917+2833W. Each of these stars is within $5''$ of another object which was closer to it in 2000 than 2015. Since the Gaia archive 2MASS cross match is not based on positions in 2015.5, but at the 2MASS epoch (approximately 2000.0), this likely explains our disagreement (Marrese et al. 2019). Out of the 5730 DR2 objects with a cataloged 2MASS identifier, 106 are objects we marked as doubles and all of these agreed with our independent cross match results. The final DR2 identifiers from our cross match can be found in Table 5.

### 4.2. Gaia Recoverability and Photometry Comparison

Out of our cross-matched targets, 915 had another DR2 star within $5''$, 350 of which we marked as having a companion within $4''$ based on Robo-AO imaging. To obtain a more accurate companion cross match, we calculated the position of each secondary Gaia DR2 star relative to the originally matched primary. We then limited our cross match to only pairing a Robo-AO multiple and a Gaia multiple if the secondary stars' positions relative to the primary differed by less than $0\rlap{.}''5$ (Fig 7).

---

[1] https://gea.esac.esa.int/archive

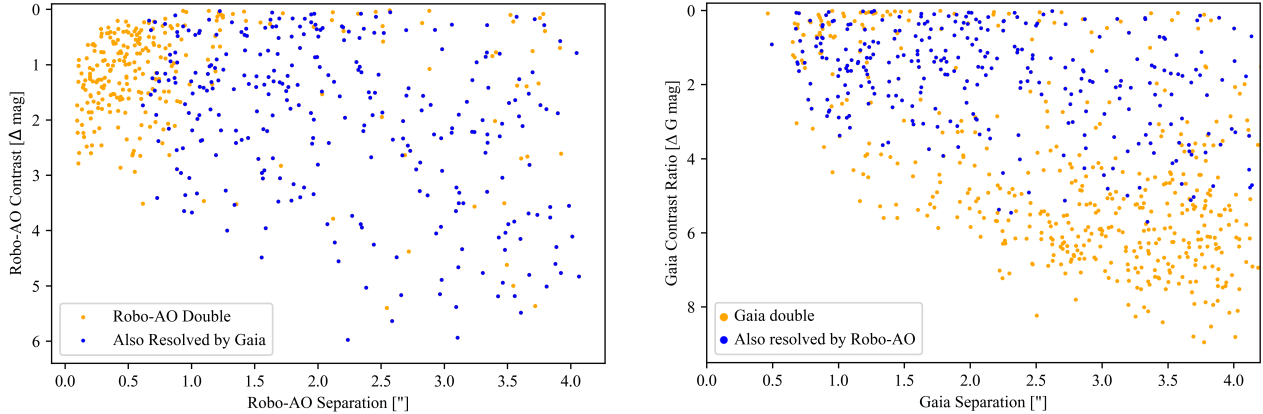

**Figure 7.** Left: the same points plotted in Fig 6. Gaia DR2 was able to resolve most of our doubles down to 1″, shown here in blue. We did not find a companion to the primary star with DR2 for the doubles shown in orange. Right: Gaia stars cross matched with out initial target list which had another star within 4″. Almost all of the near-equal brightness Gaia doubles were resolved by Robo-AO, but some were removed due to data quality. Only the pairs included in our final catalog which also had a similar arrangement to their DR2 counterpart are plotted in blue here. Gaia DR2 also resolved many doubles with contrasts beyond what can typically be detected by Robo-AO.

There are 284 of these paired systems where we have a Gaia DR2 match for both the primary and the secondary seen by Robo-AO. Using this criteria, DR2 resolved binaries down to 1″ but none within 0″.5. This agrees with a more extensive analysis of Gaia's close-binary recoverability using the Robo-AO Kepler Survey (Ziegler et al. 2018b).

For each marked Robo-AO double which was also resolved by Gaia, we determined the system's contrast and separation using the Gaia DR2 positions and $G$−band data. A comparison of the photometry between Robo-AO and Gaia can be found in Figure 8.

### 4.3. *Physical Association*

Since our observations only contain one epoch of each target, we cannot confirm which components are gravitationally bound using Robo-AO data alone. From our cross match, we obtained Gaia DR2 parallax estimates for 237 companions. Out of these, 196 had DR2 distances which agreed with the distance of their primary star to $4\sigma$. The agreement of distances in all 237 pairs are listed in Tables 3 and 4. They are expressed in units of a combined $\sigma$ from the two distance measurements, and ones with an agreement better than $4\sigma$ are likely gravitationally bound pairs.

### 4.4. *Physical Properties*

Using the Gaia DR2 parallaxes from our cross match in Section 4.1 and the determined angular separation, we estimated the projected minimum physical separation for each of our marked doubles. Here we are making the assumption that each companion is physically bound and not a background object.

To obtain mass estimates, we employ the tight empirical relationship between absolute $K$ magnitude and stellar mass (Delfosse et al. 2000; Benedict et al. 2016). We use the updated calibration of this relationship from (Mann et al. 2018), which is publicly available on `github`[2]. Given that we obtain all our $K$ photometry from 2MASS, which has an effective resolution of 2″, the $K$ magnitudes reported for most of our systems represent the blended sum of both the primary and secondary stars' flux. We could perfectly separate the two components if we had measured contrast ratios in $K$. In lieu of such measurements, we estimate $K$ contrast ratios from our measured Robo-AO $i'$ contrast ratios and the PARSEC theoretical stellar evolution models (Marigo et al. 2017). We estimate the conversion between these contrast ratios $(dK/di')$ as a function of absolute $K$, using a theoretical model fixed to 5Gyr. We interpolate along this conversion to iteratively solve for the primary and secondary $K$-band fluxes that are consistent with the blended $K$-band magnitude, the Gaia DR2 parallax of the system, and the $i'$-band contrast ratio from Robo-AO. We then estimate component masses from these individual absolute $K$ fluxes, and quote mass ratios for all companions with respect to the primary star in the system. Propagating the contrast ratio, distance, and mass-relation uncertainties results in typical mass uncertainties of 3% for the primaries and 5% for the secondaries, not accounting for any systematic uncertainties in the $i'$ to $K$ bandpass transformation.

---

[2] https://github.com/awmann/M_-M_K-

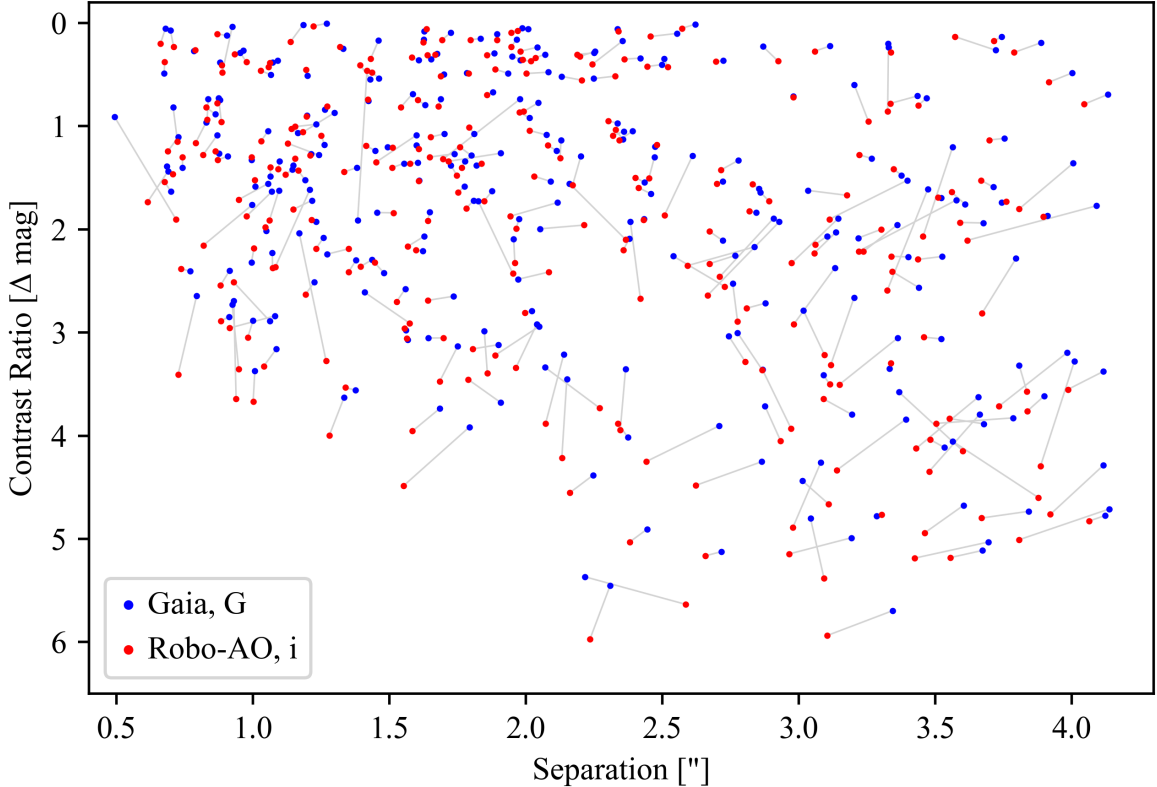

**Figure 8.** Comparing our Robo-AO photometry with Gaia DR2 from the results of our companion cross match. Plotted here are the 284 paired doubles, each pair connected by a gray line. Several pairs have a large discrepancy between the Robo-AO and Gaia separation ($> 0.''2$). This is likely because they are not physically bound pairs (due to parallax disagreement between the stars, section 4.3) and Gaia separations are determined from star positions at epoch 2015.5, while the Robo-AO observations took place 2016-2017.5.

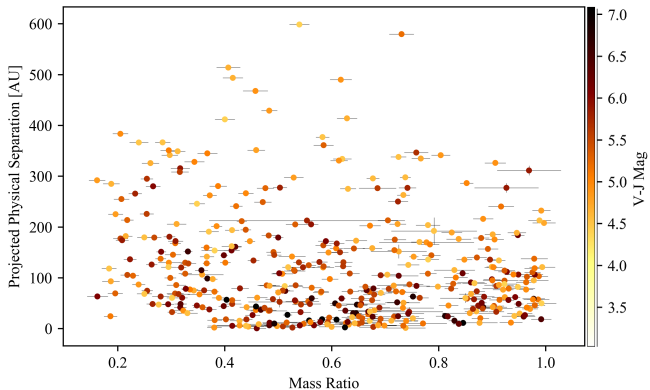

**Figure 9.** Estimated physical properties of 432 Robo-AO multiples. Projected physical separation corresponds to a minimum physical separation.

## 5. DISCUSSION

Here we discuss how target selection and observing limitations affect the sample and consider pertinence to Gaia and transiting exoplanets surveys.

### 5.1. *Survey Biases*

This sample contains several unaccounted for biases and caveats. The first of which is the survey we drew our targets from. Multiples with separations near an instrument's resolution limit may result in poor astrometry and be rejected. This could cause a bias against multiples in LSPM. Additionally, the LSPM proper motion cutoff of $0.''15$/yr includes nearly all thick-disk and halo systems, but removes many thin-disk stars. This could cause significant biases in the physical characteristics of our M dwarf sample.

We also do not have a thorough assessment of our ability to detect multiples with Robo-AO and our human-in-the-loop visual inspection, although the Gaia cross match can provide some insight. Out of the 6214 LSPM-Gaia matched stars, 682 Gaia stars had one or more stars within $4''$, ranging down to $9\Delta$ mag. Robo-AO found 335 for the same set of stars, ranging down to $6.8\Delta$ mag. Nearly all the doubles missed by Robo-AO have a contrast beyond the typical Robo-AO detection limit (Fig 7).

Additionally, we have only confirmed physical association for $\sim$200 companions using DR2 parallaxes (Section

4.3), and so our set of companions likely includes several background stars. From our Gaia cross match, out of 237 total paired doubles with parallaxes for both stars, 13 had distance discrepancies which were greater than $10\sigma$ (several of these stand out as pairs with a significant discrepancy between the Robo-AO and Gaia separation in Figure 8). This puts a rough lower-limit on the number of background stars within companions beyond $1''$. Assuming that, within this sample of nearby stars, stars within $1''$ of each other are associated, we estimate that at least 3.1% of our total doubles are not physically associated.

Our sample is affected by observation limitations, which create a bias towards wider, equal-brightness multiples, and our target selection criteria (Section 2.1). Since these stars were photometrically selected, we are more sensitive to overluminous binaries among those with a contrast $< 3\Delta$mag. The multiplicity properties within our sample are also affected by the Malmquist bias (Malmquist 1920). Selecting apparently bright stars results in a sample which is more likely to include intrinsically bright stars (higher masses) than fainter (lower mass) ones at large distances. Since doubles with close physical separations are easier to resolve in nearby systems, this would increase the multiplicity and decrease the average physical separation we observe for lower mass stars.

## 5.2. *Application of Results*

From comparing the Robo-AO multiples with multiples found in our Gaia DR2 cross match (4.2), we confirm the result in Ziegler et al. (2018b) that Gaia can resolve most companions down to $1''$ but not within that separation. Therefore Gaia has the potential for extensive multiplicity studies, though higher-angular resolution surveys are still needed to resolve close companions. This recoverability will likely improve with future releases (de Bruijne et al. 2015).

377 of our multiples do not fall within $1'$ of any systems in the Washington Double Star Catalog as of June 2019 (Mason et al. 2001). Therefore we expect the majority of these to be new discoveries. None of these would be resolved in *TESS* imaging (Ricker et al. 2014), so all of them could potentially contaminate transit observations with *TESS*. Knowing the existence and basic properties of our companions can aid in identifying and characterizing transiting exoplanets around these stars. Although none of our LSPM targets currently host confirmed exoplanets, we expect most of these stars to have planets (Dressing & Charbonneau 2015) and *TESS* is expected to find around 1,000 exoplanets around M dwarfs (Ballard 2019). Nine of our doubles appear in the Tess Habitable Zone Star Catalog (Kaltenegger et al. 2019). The LSPM collection of M dwarfs with high proper motions is a crucial target sample for *TESS*, as these tend to be the closest small stars in the sky and those around which *TESS* will be most sensitive to small transiting planets. This work presents high-resolution imaging for over 5000 of these stars. The data can be useful to the community of observers confirming and characterizing new *TESS* planets, so all Robo-AO data from this survey will be published on ExoFOP[3]. These ExoFOP data products include both annotated plots and Robo-AO observations of all multiples-star systems. They also include the contrast curve for each observation, as described in Section 2.3 and seen in Figure 3. Although these plots cannot be used for ensemble statistics due to the visual inspection we used to construct our sample, they are a useful approximation of each observation's sensitivity.

## 6. CONCLUSION

We used Robo-AO at Kitt Peak to find 553 candidate companions in 534 double systems, 8 triple systems and one quadruple system. These systems were identified from visual inspection of 5566 Robo-AO observations using `GRAM`, a graphical user interface for performing a series of visual checks and locating secondary stars. We obtained additional imaging for eleven of our targets with the NIRC2 imager on Keck II, which we used to confirm six companions. After performing an LSPM–Gaia cross match on all targets, we estimate 284 of our multiples were recovered by Gaia DR2. Out of the 237 candidate companions with DR2 distances, 196 agreed with the DR2 distance of its primary star to within $4\sigma$, implying physical association. Using these distance measurements and 2MASS absolute $K$ magnitudes we estimated the individual masses and physical separations for 432 of our multiples.

In the future, further observations of close ($< 0\rlap{.}''5$) companions are necessary for higher-significance detections and reliable photometry. Additionally, returning to reobserve potential systems not resolved by Gaia for a second epoch would confirm which companions display common proper motion with their primaries and therefore a likely physical association. For systems with wide angular separations ($\gtrsim 1''$), this could be determined through comparisons to previous multiplicity surveys and future Gaia releases. Newly discovered, close companions will require further high-angular-resolution observations. Robo-AO was removed from Kitt Peak in June 2018, and is now available on the University of

---

Hawai'i 2.2-m telescope on Maunakea, HI. This system will soon be replaced by a second generation Robo-AO system (est. mid-2020) which will deliver higher acuity and quality images that will enable the detection of much fainter companions (Baranec et al. 2018).

Although this survey is the most extensive of its kind, further statistical analysis to account for observation and selection bias needs to be done before any conclusions about multiplicity functions can be made. It would also be interesting to study multiplicity properties, particularly multiplicity as a function of mass ratio, among different mass-based subsets of our sample to explore if this M dwarf set displays common multiplicity properties or is composed of distinct populations. In a future work of this series, we plan to explore these underlying multiplicity properties within our sample.

## 7. ACKNOWLEDGEMENTS

C.L. acknowledges support from a Boettcher Foundation Educational Enrichment Grant and the Research Experience for Undergraduate program at the Institute for Astronomy, University of Hawai'i funded through NSF grant AST-1560413. She is grateful for the expertise Sean Moss contributed throughout the project. This paper is also possible through Paul Coleman's tremendous engagement and support of the IfA REU program.

C.B. acknowledges support from the Alfred P. Sloan Foundation. C.Z. is supported by a Dunlap Fellowship at the Dunlap Institute for Astronomy & Astrophysics, funded through an endowment established by the Dunlap family and the University of Toronto.

J.G.W. is supported by a grant from the John Templeton Foundation. The opinions expressed in this publication are those of the authors and do not necessarily reflect the views of the John Templeton Foundation.

The Robo-AO team thanks NSF and NOAO for making the Kitt Peak 2.1-m telescope available. Robo-AO KP is a partnership between the California Institute of Technology, the University of Hawai'i, the University of North Carolina at Chapel Hill, the Inter-University Centre for Astronomy and Astrophysics (IUCAA) at Pune, India, and the National Central University, Taiwan. The Murty family feels very happy to have added a small value to this important project. Robo-AO KP is also supported by grants from the John Templeton Foundation and the Mt. Cuba Astronomical Foundation. The Robo-AO instrument was developed with support from the National Science Foundation under grants AST-0906060, AST-0960343, and AST-1207891, the Mt. Cuba Astronomical Foundation, and by a gift from Samuel Oschin. These data are based on observations at Kitt Peak National Observatory, National Optical Astronomy Observatory (NOAO Prop. ID: 15B-3001), which is operated by the Association of Universities for Research in Astronomy (AURA) under cooperative agreement with the National Science Foundation.

Some of the data presented herein were obtained at the W. M. Keck Observatory, which is operated as a scientific partnership among the California Institute of Technology, the University of California and the National Aeronautics and Space Administration. The Observatory was made possible by the generous financial support of the W. M. Keck Foundation.

The authors wish to recognize and acknowledge the very significant cultural role and reverence that the summit of Maunakea has always had within the indigenous Hawaiian community. We are most fortunate to have the opportunity to conduct observations from this mountain.

Some of the data presented in this paper were obtained from the Mikulski Archive for Space Telescopes (MAST). STScI is operated by the Association of Universities for Research in Astronomy, Inc., under NASA contract NAS5-26555.

This research has made use of the SIMBAD database, operated at CDS, Strasbourg, France.

This research has made use of the Exoplanet Follow-up Observation Program website, which is operated by the California Institute of Technology, under contract with the National Aeronautics and Space Administration under the Exoplanet Exploration Program.

This research has made use of the Washington Double Star Catalog maintained at the U.S. Naval Observatory.

This work has made use of data from the European Space Agency (ESA) mission *Gaia* (https://www.cosmos.esa.int/gaia), processed by the *Gaia* Data Processing and Analysis Consortium (DPAC, https://www.cosmos.esa.int/web/gaia/dpac/consortium). Funding for the DPAC has been provided by national institutions, in particular the institutions participating in the *Gaia* Multilateral Agreement.

This publication makes use of data products from the Two Micron All Sky Survey, which is a joint project of the University of Massachusetts and the Infrared Processing and Analysis Center/California Institute of Technology, funded by the National Aeronautics and Space Administration and the National Science Foundation.

*Facilities:* KPNO:2.1m (Robo-AO), Keck:II (NIRC2-NGS)

APPENDIX

**Table 2. Keck/NIRC2 Follow-up Observations and Results**

| LSPM ID | Filter | Exposure | # images | Proj Sep. | Angle | Contrast | Contrast |
|---------|--------|----------|----------|-----------|-------|----------|----------|
|         |        | s        |          | ''        | deg   | Δmag     | error    |
| J1504+2928 | Kp  | 120      | 1        |           |       |          |          |
| J1558+4927 | Kp  | 120      | 1        |           |       |          |          |
| J1606+0823 | Kp  | 167      | 4        | 2.445     | 69    | 1.77     | 0.02     |
|            |     |          |          | 3.04      | 69    | 1.95     | 0.02     |
|            | J   | 167      | 3        | 2.454     | 69    | 1.8      | 0.02     |
|            |     |          |          | 3.053     | 69    | 1.97     | 0.02     |
| J1648+1038 | Kp  | 167      | 5        | 0.135     | 305   | 0.2      | 0.08     |
|            | J   | 167      | 3        | 0.153     | 306   | 0.48     | 0.15     |
| J1703+3211 | Kp  | 167      | 4        | 1.464     | 206   | 1.54     | 0.03     |
|            | J   | 167      | 3        | 1.487     | 208   | 1.6      | 0.09     |
| J1825+4040 | Kp  | 60       | 2        |           |       |          |          |
| J1922+2146 | Kp  | 60       | 4        |           |       |          |          |
|            | J   | 60       | 2        |           |       |          |          |
| J2127+5505 | Kp  | 60       | 1        |           |       |          |          |
| J2122+3025 | Kp  | 60       | 4        | 0.253     | 69    | 0.08     | 0.03     |
|            | J   | 60       | 4        | 0.254     | 7     | 0.05     | 0.03     |
| J2210+4417 | K   | 167      | 4        | 0.526     | 288   | 1.52     | 0.02     |
|            | J   | 167      | 2        | 0.528     | 288   | 1.59     | 0.02     |
| J2251+4921 | J   | 167      | 4        | 0.245     | 271   | 0.25     | 0.02     |
|            | K   | 167      | 1        | 0.245     | 271   | 0.25     | 0.05     |

NOTE—Data was taken 2017 August 3 with the NIRC2 camera on Keck II. Companion properties are shown here in the case that one was detected. J1606+0823, a triple system, has analysis for two companions in each set of images.

**Table 3. Robo-AO Higher-Order Multiples**

| Target | Companion | Proj Sep. | Angle | Contrast | Detection | Phys Sep | Mass Ratio | DR2 distance | WDS ID |
|--------|-----------|-----------|-------|----------|-----------|----------|------------|--------------|--------|
|        |           | ''        | deg   | Δmag     | significance | AU    | ratio      | agreement[σ] |        |
| J1047+2007E | B    | 3.78±0.07 | 239.0±1 | 0.08±0.01 | 4.0    | 137.6±0.8 | 0.98±0.03 | -          | J10476+2008 |
|             | C    | 0.26±0.07 | 188.0±1 | 1.7±0.09  | 94.0   | 9.5±0.8   | 0.55±0.02 | -          | -          |
| J1417+7031S | B    | 3.59±0.07 | 345.0±1 | 1.26±0.01 | 86.0   | 213.2±0.1 | 0.71±0.02 | 4.3        | J14180+7032 |
|             | C    | 3.46±0.07 | 338.0±1 | 2.07±0.02 | 11.0   | 205.2±0.1 | 0.53±0.02 | -          | -          |
| J1606+0823 | B     | 2.36±0.07 | 291.0±1 | 2.2±0.01  | 1.0    | -         | -         | -          | J16067+0823 |
|            | C     | 2.92±0.07 | 290.0±1 | 2.26±0.01 | 1.0    | -         | -         | -          | -          |

*Table 3 continued*

**Table 3** *(continued)*

| Target | Companion | Proj Sep. | Angle | Contrast | Detection | Phys Sep | Mass Ratio | DR2 distance | WDS ID |
|---|---|---|---|---|---|---|---|---|---|
| | | $''$ | deg | $\Delta$mag | significance | AU | ratio | agreement[$\sigma$] | |
| J1607+2955 | B | 0.7±0.07 | 189.0±3 | 0.47±0.03 | 13.0 | - | - | - | - |
| | C | 3.54±0.07 | 234.0±1 | 5.19±0.04 | 38.0 | - | - | - | - |
| J1727+2208 | B | 2.82±0.07 | 251.0±1 | 0.24±0.02 | 64.0 | 311.4±9.4 | 0.97±0.06 | - | - |
| | C | 2.51±0.07 | 256.0±1 | 0.58±0.02 | 45.0 | 277.2±9.4 | 0.93±0.06 | - | - |
| | D | 0.3±0.07 | 351.0±7 | 1.42±0.11 | 6.0 | 33.1±9.4 | 0.74±0.06 | - | - |
| J1922+2146 | B | 3.66±0.07 | 277.0±1 | 3.69±0.03 | 30.0 | 179.3±0.1 | 0.27±0.01 | - | - |
| | C | 2.16±0.07 | 128.0±1 | 4.55±0.05 | 79.0 | 105.7±0.1 | 0.22±0.01 | - | - |
| J1948+3250 | B | 1.68±0.07 | 321.0±1 | 3.47±0.08 | 11.0 | 99.9±0.1 | 0.3±0.01 | 1.1 | - |
| | C | 3.61±0.07 | 88.0±1 | 5.48±0.44 | 5.0 | 213.8±0.1 | 0.22±0.01 | - | - |
| J1958+3217 | B | 2.27±0.07 | 58.0±1 | 3.73±0.07 | 18.0 | 72.8±0.1 | 0.38±0.01 | 4.5 | - |
| | C | 4.01±0.07 | 343.0±1 | 4.11±0.1 | 19.0 | 128.6±0.1 | 0.37±0.01 | - | - |
| J2141+2741 | B | 0.26±0.07 | 128.0±1 | 1.51±0.1 | 4.0 | 6.5±0.3 | 0.57±0.01 | - | - |
| | C | 2.56±0.07 | 303.0±1 | 6.77±0.08 | 10.0 | 63.4±0.3 | 0.16±0.02 | - | - |

Note—Eight potential triple systems and one quadruple. The first column is the primary star's LSPM identifier and the following set of six columns are properties determined through Robo-AO. Physical separation and the agreement between the distances of the two components were determined through a Gaia DR2 cross match. When applicable, the likely Washington Double Star Catalog identifier is given. For original references to previously discovered systems, see the Washington Double Star Catalog at https://go.nasa.gov/2zTEV2W

**Table 4. Robo-AO Double Stars**

| Target | Proj Sep. | Angle | Contrast | Detection | False | Phys Sep | Mass | DR2 distance | WDS ID |
|---|---|---|---|---|---|---|---|---|---|
| | $''$ | deg | $\Delta$mag | significance | triple[a] | AU | ratio | agreement [$\sigma$] | |
| J0006+2736 | 1.55±0.07 | 215.0±2 | 0.24±0.04 | 24.0 | - | 134.3±0.7 | 0.91±0.01 | - | - |
| J0008+2821 | 1.03±0.07 | 117.0±1 | 0.46±0.03 | 41.0 | x | 91.0±2.4 | 0.86±0.03 | 1.0 | - |
| J0010+3646 | 1.41±0.07 | 189.0±2 | 0.04±0.01 | 87.0 | - | 103.2±0.4 | 0.99±0.02 | - | - |
| J0014+2822N | 1.98±0.07 | 130.0±1 | 0.27±0.02 | 70.0 | - | 121.6±0.3 | 0.9±0.01 | 0.8 | - |
| J0015+3642N | 2.51±0.07 | 269.0±359 | 0.3±0.02 | 39.0 | - | 175.6±0.3 | 0.9±0.02 | - | J00152+3642 |
| J0016+3000 | 0.88±0.07 | 174.0±2 | 2.54±0.03 | 15.0 | - | 36.7±0.3 | 0.46±0.02 | - | - |
| J0020+4248 | 1.76±0.07 | 286.0±1 | 1.2±0.01 | 78.0 | - | 125.4±0.2 | 0.74±0.02 | 1.7 | - |
| J0021+2923 | 0.31±0.07 | 358.0±7 | 1.23±0.05 | 8.0 | - | 23.5±4.6 | 0.72±0.05 | - | - |
| J0059+3752 | 0.93±0.07 | 322.0±1 | 0.3±0.05 | 37.0 | x | 42.3±0.2 | 0.88±0.01 | 1.9 | J00591+3753 |
| J0103+3140 | 0.2±0.07 | 148.0±4 | 1.57±0.16 | - | - | - | - | - | - |
| J0106+3336S | 3.34±0.07 | 16.0±1 | 0.28±0.03 | 38.0 | - | 135.8±0.1 | 0.9±0.01 | 0.1 | - |
| J0107+3326 | 3.09±0.07 | 141.0±1 | 3.64±0.13 | 12.0 | - | 513.9±1.2 | 0.41±0.02 | 0.6 | - |
| J0123+3559 | 1.68±0.07 | 32.0±2 | 0.81±0.02 | 51.0 | - | 63.2±0.2 | 0.75±0.01 | 0.3 | J01230+3600 |
| J0142+3702 | 0.29±0.07 | 142.0±2 | 1.27±0.05 | 7.0 | - | - | - | - | - |
| J0157+3737 | 3.33±0.07 | 61.0±1 | 0.85±0.02 | 80.0 | - | 277.1±0.5 | 0.74±0.02 | 17.4 | - |
| J0220+3320 | 1.52±0.07 | 337.0±1 | 1.84±0.04 | 79.0 | - | 80.5±0.2 | 0.56±0.02 | 0.5 | J02208+3321 |
| J0221+3653 | 1.2±0.07 | 341.0±1 | 0.2±0.01 | 71.0 | - | - | - | - | - |
| J0236+3204W | 2.03±0.07 | 58.0±1 | 0.34±0.01 | 67.0 | - | 39.7±0.1 | 0.88±0.01 | 0.3 | - |
| J0259+3636 | 1.81±0.07 | 5.0±1 | 3.16±0.03 | 32.0 | - | 41.9±0.1 | 0.32±0.01 | 1.2 | J02592+3637 |
| J0259+3855 | 0.88±0.07 | 17.0±3 | 1.12±0.02 | 29.0 | - | - | - | - | J02598+3856 |

Note—534 potential companions with properties determined through Robo-AO. As in Table 3, the third set of columns display properties determined through cross matches with existing catalogs 4.4. When applicable, the likely Washington Double Star Catalog identifier is also given. The first 20 lines are displayed here. For original references to previously discovered systems, see the Washington Double Star Catalog at https://go.nasa.gov/2zTEV2W

[a] The position angle of systems marked as a false triple may be offset by 180°.

**Table 5.** Cross-Match between LSPM and Gaia DR2

| LSPM ID | Gaia DR2 ID | Data Observed YYYYMMDD | # Robo-AO Companions | # Gaia Companions |
|---------|-------------|------------------------|----------------------|-------------------|
| J0000+2546 | 2853247281263713024 | 20161016 | 0 | 0 |
| J0000+3149 | 2873735615295675776 | 20161009 | 0 | 0 |
| J0000+3758 | 2880533208495595136 | 20161011 | 0 | 0 |
| J0000+3951 | 2881861246743522176 | 20161012 | 0 | 0 |
| J0001+2559 | 2853263121103114368 | 20161015 | 0 | 0 |
| J0001+3147 | 2873754272633700480 | 20161011 | 0 | 0 |
| J0001+3828 | 2880971840623200896 | 20161011 | 0 | 0 |
| J0001+4251 | 384465994663239808 | 20161014 | 0 | 0 |
| J0002+2430 | 2849968228352159232 | 20161019 | 0 | 0 |
| J0003+3617 | 2877184645833112064 | 20161010 | 0 | 0 |
| J0004+2456 | 2850097554111353088 | 20161018 | 0 | 0 |
| J0005+4129 | 384052750089843328 | 20161012 | 0 | 0 |
| J0005+4547 | 386655019234959872 | 20161015 | 0 | 0 |
| J0006+2552 | 2850587253398091776 | 20161018 | 0 | 0 |
| J0006+2736 | 2854029549426321664 | 20161015 | 1 | 1 |
| J0006+4000 | 383162661067592192 | 20161012 | 0 | 0 |
| J0006+4342 | 384739773058873088 | 20161015 | 0 | 0 |
| J0007+4101 | 383981178754905984 | 20161014 | 0 | 0 |
| J0008+2739 | 2853856101467493760 | 20161015 | 0 | 0 |
| J0008+2821 | 2860084010205910400 | 20161011 | 1 | 1 |

NOTE—Every Robo-AO target observed with the results from the Gaia DR2 cross match (Section 4.1). The first 20 lines are displayed here.